\documentclass[useAMS,usenatbib]{mn2e}
\usepackage{graphicx}
\usepackage{graphics}
\usepackage{epstopdf}
\usepackage{float}
\usepackage{placeins}
\usepackage{bm}
\usepackage{epsf}
\usepackage{url}
\usepackage{epsfig}
\usepackage{natbib}
\usepackage{amsmath}

\title[Scaling Relations in the Network of Voids]{Scaling Relations in the Network of Voids: Implications for Local Universe Dynamics and Inferring the Metric of Space}
\author[Aragon-Calvo M.A. et al.]{M.A. Aragon-Calvo$^{1}$\thanks{E-mail:maragon@astro.unam.mx} \\
$^{1}$Instituto de Astronom\'{i}a, UNAM, Apdo. Postal 106, Ensenada 22800, B.C., M\'{e}xico\\}

\begin{document}


\pagerange{\pageref{firstpage}--\pageref{lastpage}} \pubyear{2002}
\maketitle
\label{firstpage}

\begin{abstract}

The large-scale distribution of matter in the universe forms a network of clusters, filaments and walls enclosing large empty voids. Voids in turn can be described as a cellular system in which voids/cells define dynamically distinct regions. Cellular systems arising from a variety of physical and biological processes have been observed to closely follow scaling laws relating their geometry, topology and dynamics. These scaling laws have never been studied for cosmological voids, the largest known cellular system. 
Using a cosmological N-body simulation we present a study of the scaling relations of the network of voids, extending their validity by over 30 orders of magnitude in scale with respect to other known cellular systems. 

Scaling relations allow us to make indirect measurements of the dynamical state of voids from their geometry and topology. Using our results we interpret the ``local velocity anomaly"  observed in the Leo Spur as the result of a collapsing void in our cosmic backyard. 

Moreover, the geometry and connectivity of voids directly depends on the curvature of space. Here we propose scaling relations as an independent and novel measure of the metric of space and discuss their use in future galaxy surveys.

\end{abstract}

\begin{figure}
\includegraphics[width=0.49\textwidth]{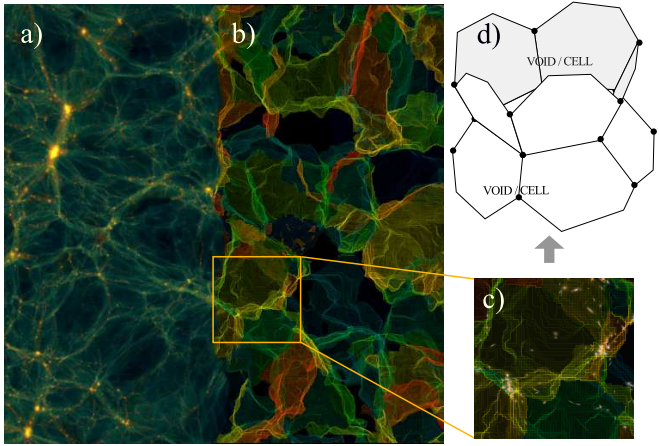}
\caption{The Cosmic Web delineated by voids as a cellular system. a) dark matter density field, colors indicate local density. b) The corresponding void network computed using the Spine method. We show the walls delineating voids as semitransparent membranes in the volume rendering. c) zoom region showing galaxies (indicating the position of dark matter haloes) superimposed on the void network. d) Toy model of the zoomed region. The voids, walls, filaments and clusters delineated by luminous galaxies correspond to cells, faces, edges and nodes of a cellular system.}
 \label{fig:cosmic_web_cellular_system}
\end{figure}

\section{Introduction}

The Cosmic Web is a dynamic system evolving under the action of gravity towards the configuration of minimum energy. Tiny density fluctuations in the primordial density field grew accentuated by gravity to form the galaxies and their associations we observe today. The distribution of galaxies forms an interconnected network of flat walls, elongated filaments and compact clusters of galaxies delineating vast empty void cells. Voids, walls, filaments and clusters have been since their discovery associated to cells, faces, edges and vertices of a cellular system \citep{Einasto80, Icke84, Icke87, Weygaert89,Dubinski93, Weygaert94,Neyrinck14} (see Fig. 1). The geometry and topology of the network of voids has a tantalizing similarity to other cellular systems observed in nature. Even its evolution resembles the coarsening of soap foam where large bubbles grow by the collapse of adjacent smaller bubbles \citep{Plateau1873}. This is similar  to the way voids grow by their own expansion and by the collapse of smaller voids in their periphery \citep{Sheth04}. 

Cellular systems arise from a disparate variety of physical and biological systems as the result of common spatial and topological processes. Cellular systems have been observed to follow a set of scaling relations relating the connectivity and geometry of their cells over a wide range of scales \citep{Rivier93}.  In this work we will study three well-known scaling laws:  

\begin{itemize}
\item The Lewis law  \citep{Lewis28}, gives a linear relation between the area of a given cell and its number of neighbors, or its degree $n$ as :

\begin{equation}
	n = c_1 + c_2 A.
\end{equation}

\noindent Where $A$ is the area of the central cell and $c_1, c_2$ are constants. 

\item The Aboav-Weaire  law \citep{Aboav70, Weaire99}, describes the inverse relation between the degree of a cell and the degree of its adjacent cells: the largest the degree of a cell, the smallest the degree of its neighbours. This relation is usually expressed as :

\begin{equation}\label{eq:aboav}
	m_n = a + b/n, 
\end{equation}
	
\noindent where $m_n$ is the degree of the adjacent cells. A simpler alternate form of the Aboav-Weaire relation is:

\begin{equation} \label{eq:aboav_linear}
 n \cdot m_n = (<n> - \mu) + (<n>-1) \cdot n.
\end{equation}

\noindent such that the product $n \cdot m_n$ is linearly dependent on $n$. 

\item The von Neumann law \citep{Neumann52} describes the rate of change of the area of cells  with their degree. It was originally analytically derived for the 2D case. In its simplest form it is expressed as:

\begin{equation}
	dA/dt = k(n-6),
\end{equation}

\noindent where $dA/dt$ is the rate of change of the area of the cell and $k$ is a constant. This remarkable relation is purely topological, relating the dynamical state of a cell only with its connectivity. Note that in the 2D case the stable configuration for a cellular system is the hexagonal lattice (n=6). The long-sought extension of the von Neumann relation to three dimensions was relatively recently derived by \citet{MacPherson07} (see also \citet{Lazar14} for a recent study of scaling relations on Voronoi systems).

\end{itemize}

\subsection{This paper}

The scaling relations described above have been measured in cellular systems arising from different physical processes acting over a wide range of scales. In this work we measure these scaling relations on the network of voids identified from an N-body simulation.  We then use our results to interpret the dynamics of the local Universe and discuss their use to directly prove the metric of space at large scales.

%
\section{Computer simulations and void catalogues}

The results presented in this work are based on a N-body computer simulation containing $128^3$ dark matter particles inside a box of 100  $h^{-1}$  Mpc side with the standard $\Lambda$CDM cosmology: $\Omega_m=0.3$, $\Omega_{\Lambda}=0.7$, $h=0.73$, $\sigma_8=0.8$. Starting at $z=80$ we evolved the box to the present time using the N-body code {\small GADGET-2} \citep{Springel05} and stored 32 snapshots in logarithmic intervals of the expansion factor starting at $z=10$. While the particle number may seem somewhat low it is sufficient for our purposes since the smallest voids we are interested in are larger than a few Mpc in radius. The low particle number acts as a low-pass filter, removing structures that could result in unwanted over-segmentation in the watershed method used to identify voids.

\subsection{Void identification}\label{sec:void_catalog}
From the particle distribution at each of the 32 snapshots we computed a continuous density field on a regular grid of $512^3$ voxels using the openMP implementation of the Lagrangian Sheet density estimation method \citep{Abel12, Shandarin12}  as described in \citet{Aragon14} (see Data Availability, Sec. \ref{sec:data_availbility}). The Lagrangian nature of this density estimation and interpolation method allows us to compute accurate densities at very early times and also at latter times inside voids where the particle arrangement is still close to a regular grid. From the density fields we identified voids using the floating-point implementation of the watershed void finder \citep{Platen07} included in the Spine pipeline \citep{Aragon10b}.

\subsection{Tracking voids in time}

We follow the evolution of individual voids by constructing their ``Eulerian" merging tree in a similar fashion as done with
haloes in N-body simulations but instead of tracking particles we track voxels. 
Two voids identified in adjacent snapshots are linked if they share common voxels as:

\begin{equation}
V_i \cap V_{i+1} > 0
\end{equation}

\noindent where $V_i$ and $V_{i+1}$ are two voids in adjacent snapshots $i$ and $i+1$. A void in snapshot $i$
is linked to its ``progenitor" in snapshot $i+1$ if it shares with it more than 75 percent of its volume. Using this
procedure we created for each void in the simulation its \textit{progenitor line}. This approach is the same as described in \citet{Aragon10b} but performing the linking across time (adjacent snapshots) instead of scale (hierarchical space). 

\subsection{Void degree and void graph}

We compute the degree of each void from the incomplete watershed regions i.e the regions without their boundaries (see \citet{Aragon10b}).  For each void we construct a mask with value 1 for the void voxels and 0 outside. This mask is then dilated by one voxel using the dilation morphology operator:

\begin{equation}
M_i = V_i \otimes K = \bigcup_{k \in K} \; V_k,
\end{equation}

\noindent where $M_i$ is the mask around void $i$, $V_i$ is the set of voxels defining the void and $K$ is the structuring 
kernel composed by a grid of $3\times3\times3$ voxels. The adjacent voids are then identified by intersecting the mask
with the watershed regions:

\begin{equation}
A_j = M_i \cap W
\end{equation}

\noindent where $A_j$ contains the list of adjacent voids and $W$ is the set of all watershed regions.
The degree of the central void is the number of unique void labes in the set $A_j$.

From the list of adjacent voids we created a \textit{void-graph} with nodes corresponding to void centers and edges joining adjacent voids (this \textit{void-graph}  forms a triangulation which is the dual of the \textit{cosmic web-graph}). Note that in in the void-graph the edges correspond, by construction,  to walls.

\begin{figure}
  \centering
  \includegraphics[width=0.45\textwidth,angle=0.0]{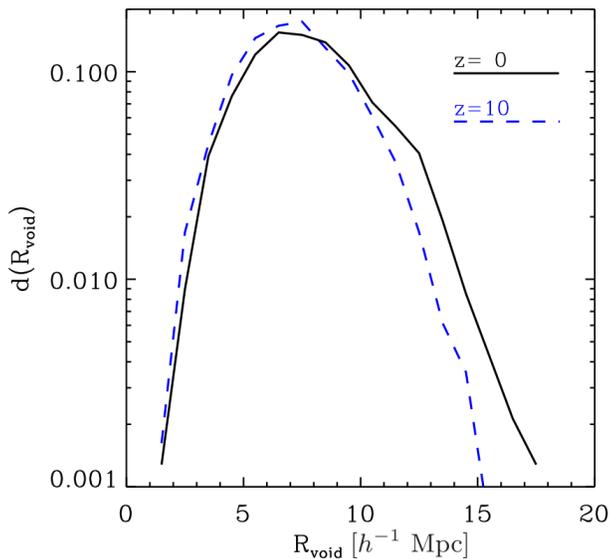}
  \caption{Distribution of void equivalent radius $R_{\textrm{\tiny{void}}}$ at $z=0$ (solid line) and $z=10$ (dashed line). }
  \label{fig:radius_distribution}
\end{figure}

%
\subsection{Void radius and degree}

Figure \ref{fig:radius_distribution} shows the distribution of void radius at redshifts $z=10$ and $z=0$. We define the
effective radius of a void as the radius of a sphere with the same volume as the void:

\begin{equation}
R_{\textrm{\tiny{void}}} =  ( 3 / (4 \pi) V_{\textrm{\tiny{void}}} )^{1/3}.
\end{equation}

\noindent Where $V_{\textrm{\tiny{void}}}$ is the sum of all voxels defining the void. Given the finite size of the
voxels there is a discreetness effect that affects the computation of the radius for voids with radius smaller than a few times the voxel size.
The distribution of void radius is approximately symmetrical at early times and skews towards larger radius at the present time, reflecting the growth of large voids and the collapse of smaller ones (see also \citet{Platen08}).

\begin{figure}
  \centering
  \includegraphics[width=0.45\textwidth,angle=0.0]{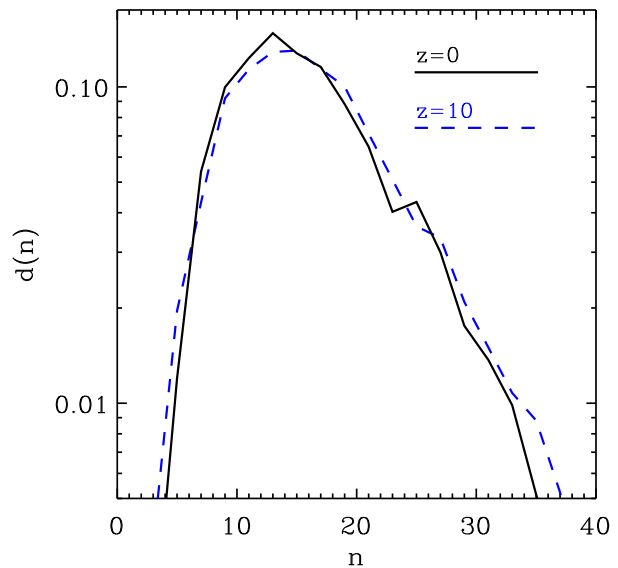}
  \caption{Distribution of void degree as a function of its radius $R_{\textrm{\tiny{void}}}$ at $z=0$ (solid line) and $z=10$ (dashed line). }
  \label{fig:degree_distribution}
\end{figure}

\begin{figure}
  \centering
  \includegraphics[width=0.45\textwidth,angle=0.0]{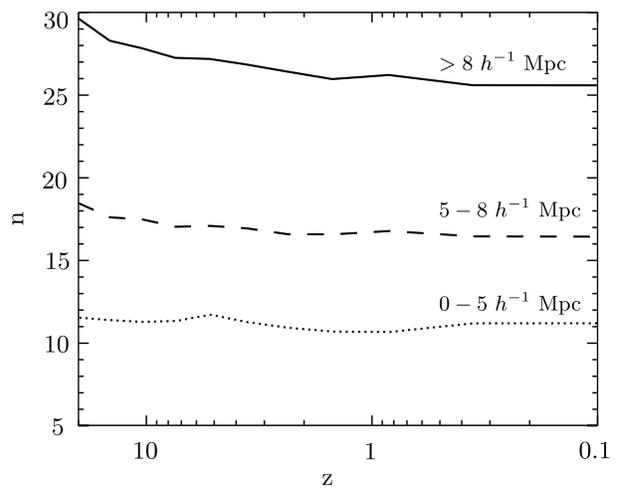}
  \caption{Mean void degree as a function of redshift for three void radius intervals (see text for details). }
  \label{fig:degree_z_radius}
\end{figure}

\begin{figure*}
  \centering
  \includegraphics[width=0.99\textwidth]{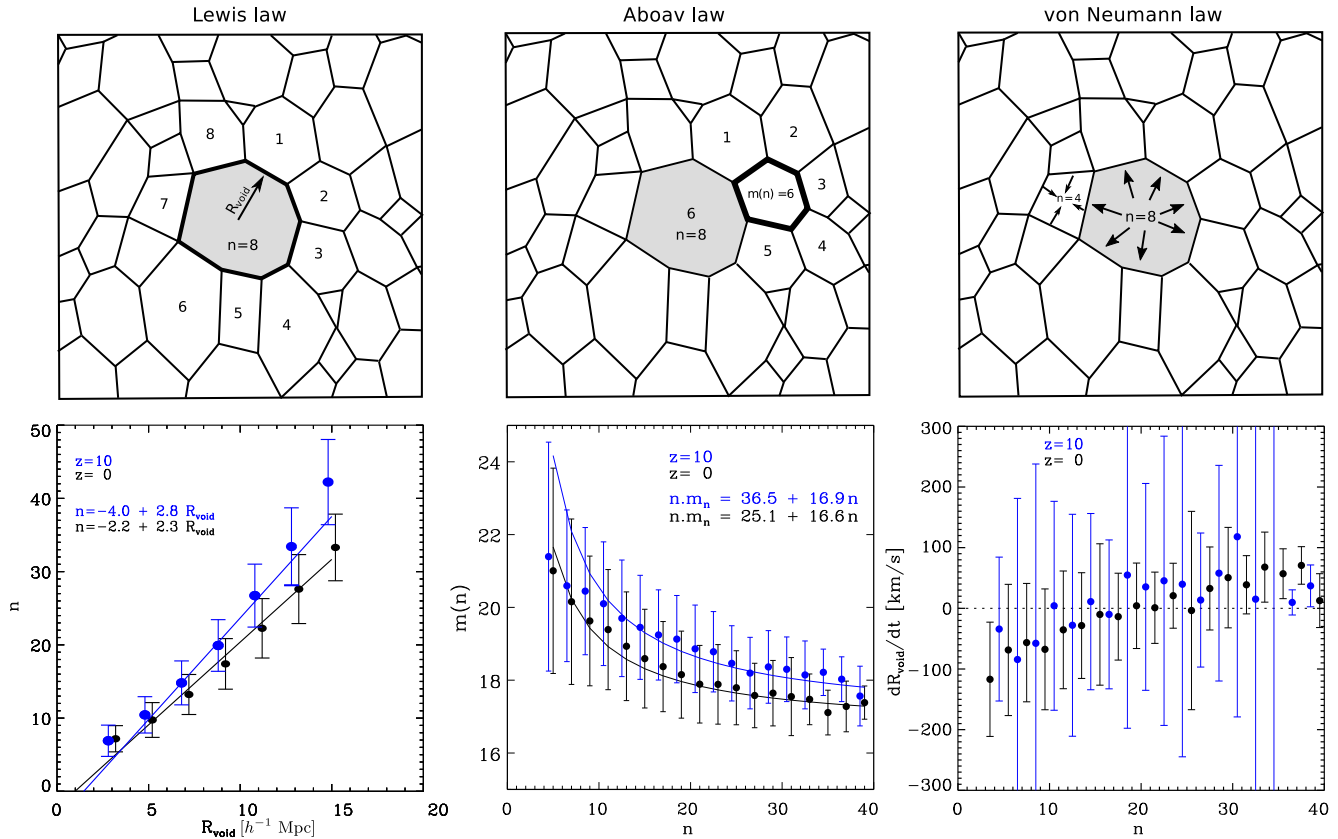}
  \caption{Scaling relations in the network of voids computed at $z=10$ and $z=0$. \textbf{Left panels}: Void degree as a function of  radius (Lewis law). $R_{\textrm{\tiny{void}}} $ is the effective radius of the void (see text for details) and $n$ is its degree. Solid lines correspond to a linear fit with parameters shown on the top-left corner. The top panel shows a central void (thick black contour) with $n=8$.
  \textbf{Center panels}: Adjacent void degree $m(n)$ as a function of the central void's degree (Aboav's law). The mean inside each bin is indicated by large dots, solid lines correspond to a linearized fit (see text for details) with parameters indicated on the top-right.  The top panel shows the central void with $n=8$ and one of its adjacent voids (thick black contour) with degree $m(n)=6$.
  \textbf{Right panels}: Rate of change of void radius $dR_{\textrm{\tiny{void}}} /dt$ with respect to their degree. The mean inside each bin is indicated by large dots, the dotted line indicates the point of equilibrium. Error bars in all plots show the dispersion inside each bin.}
  \label{fig:scaling_relations}
\end{figure*} 

%
\subsection{Mean void degree}\label{sec:void_degree}

Figure \ref{fig:degree_distribution} shows the distribution of voids degree as a function of radius  at $z=10$ and $z=0$. 
The peak of the distribution is around $n \sim 13$ which is similar to the value expected for a Voronoi 
distribution where $n = 48/35 \pi^2 + 2 = 15.53$ \citep{Sullivan99}. 

Figure \ref{fig:degree_z_radius} shows the mean degree for three bins in radius as a function of redshift. There is an obvious dependency between void size and its degree, as expected since large voids have a larger surface area shared with a larger number of voids.

For comparison we computed the mean degree from a Voronoi distribution where the density field is replaced by the distance field which increases with its distance from the Voronoi centers. We identified the Voronoi cells
following the same approach used to identify voids in Section \ref{sec:void_catalog} and obtained 
a value of $n \sim 16$. \citet{Weygaert94} obtained similar values for the number of faces of a Voronoi distribution computed directly
from the Voronoi tessellation of random, correlated and anti-correlated ``seed points".

%
\section{Results}

%
\subsection{Lewis relation}

Figure \ref{fig:scaling_relations} (left panel) shows the Lewis scaling relation computed from the void network at two different times. 
We found that, consistent with the Lewis law reported elsewhere, large voids have on average a higher degree than smaller voids (left panel).  At the present time the relation is linear while at high redshift there is a small departure from linearity at the extreme small and large voids.  The peak of the void radius distribution (Fig. \ref{fig:radius_distribution}) is $R_{\textrm{\tiny{void}}}  \sim 7 h^{-1}$ Mpc at $z = 0$ corresponding to $n \sim 14$ in Fig. \ref{fig:scaling_relations} left panel. This value is close to the mean number of neighbors computed for Voronoi distributions $n_{\textrm{\tiny{Voro}}} =  15.54$ as discussed in Sec. \ref{sec:void_degree} (see also \citet{Meijering53} and \citet{Weygaert94} for an extensive study of Voronoi tessellations). 
Interestingly, predictions for a void network arising from a phase transition give $n = 13.4$ \citep{Laix99} also within our measured values. 
At early times there is a stronger dependence of $n$ on $R_{\textrm{\tiny{void}}} $ than at the present time. In fact at $z=10$ the mean degree of voids is closer to the Voronoi case. This is consistent (at least qualitatively) with the study by \citet{Weygaert94} who found that the mean degree of Voronoi cells from a Poisson distribution decreases as the Poisson seeds become correlated, this being a crude approximation to the correlating effect of gravity on an initial Poisson distribution of Voronoi seeds.

%
\subsection{Aboav relation}

Figure \ref{fig:scaling_relations} (center panel) shows the Aboav scaling relation measured as the average neighbor number $m$ of the neighbors cells as function of the number of neighbors $n$ of the central cell. For simplicity we plot the direct relation (eq. \ref{eq:aboav}) but perform a fit of the linearized form (eq. \ref{eq:aboav_linear}).

Voids follow the Aboav relation at all times. At high redshifts and low $n$ a decrease in the slope can be observed, although we do not see the full downturn measured in Voronoi distributions \citep{Hilhorst06, Lazar14}. The higher values of $m$ at early times suggest a more uniform initial distribution of void sizes.

\begin{figure}
  \centering
  \includegraphics[width=0.45\textwidth]{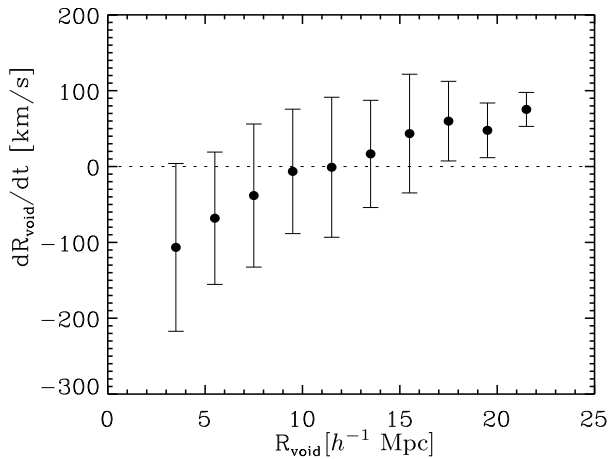}
  \caption{Rate of change of void radius $dR_{\textrm{\tiny{void}}} /dt$ as a function of their radius.  $dR_{\textrm{\tiny{void}}} /dt$ corresponds to the mean velocity of the particles at the boundaries of the void measured from the void's center. The bars indicate the dispersion inside each bin.}
  \label{fig:von_neumann}
\end{figure}

%
\subsection{von Neumann relation: topology and dynamics}

Figure \ref{fig:scaling_relations} (right panel) shows the von Neumann relation measured as the rate of change of void radius with respect to their degree. While there is a clear relation at both $z=10$ and $z=0$ the dispersion increases by a factor of two at low redshift possibly reflecting non-linear processes in the void evolution. Voids with a lower number of adjacent voids than the measured critical value $n_{\textrm{\tiny{crit}}} $ (where $dR/dt = 0$) tend to collapse. This critical degree is $n_{\textrm{\tiny{crit}}}  \sim 18$ at the present time. 

Figure \ref{fig:von_neumann} shows $dR_{\textrm{\tiny{void}}} /dt$ as a function of their equivalent radius. The intersection point at $n \sim 20$ in Fig. \ref{fig:scaling_relations} corresponds to $R_{\textrm{\tiny{crit}}}  \sim 9 h^{-1}$ Mpc. Note that this can be directly obtained from the Lewis relation in Fig. \ref{fig:scaling_relations} (left panel). Voids smaller than $R_{\textrm{\tiny{crit}}}$ tend to collapse. It is interesting to compare the von Neumann relation with the distribution of void sizes (Fig. \ref{fig:radius_distribution}) where the distributions at $z=10$ and $z=0$ intersect at $R_{\textrm{\tiny{void}}} \sim 8$ Mpc $h^{-1}$, suggesting that voids smaller than this radius collapse and larger voids expand.

%
\section{Discussion}

\begin{figure}
\includegraphics[width=0.49\textwidth]{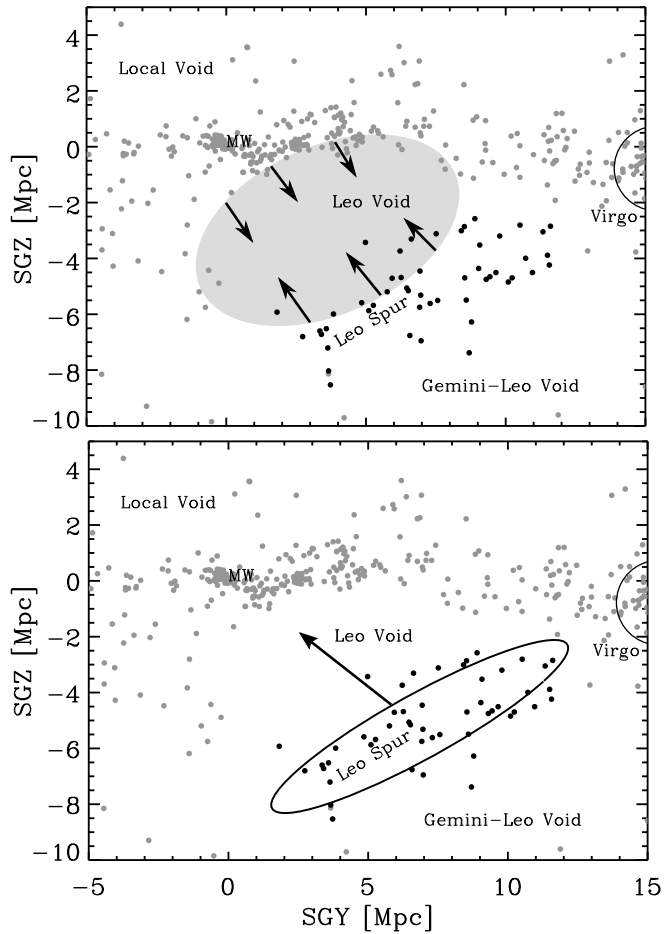}
\caption{Distribution of galaxies in a thin slice over the supergalactic X axis showing the geometry and dynamics in the region around the Leo Void. The Milky Way is located at the origin of the coordinate system and the Local Sheet is seen edge-on lying on the $SGZ =0$ plane. Top panel: proposed collapse of the Leo Void (gray ellipse). The arrows indicate the direction of collapse.  Bottom panel: resulting observed peculiar velocity ($ \sim 200$ km/s) of the Leo Spur (black ellipse).}
 \label{fig:collapsing_void}
\end{figure}

\subsection{The origin of the Local Sheet and a collapsing void in our cosmic backyard}

We propose that the Local Sheet in which the Milky Way is located may be the result of the collapse of a small void. In this scenario the galaxies in the plane of the Local Sheet were once located in the periphery of a small void of radius $\sim 4$ Mpc \citep{Aragon11,McCall14} that collapsed into its current flat shape. Figure \ref{fig:collapsing_void} shows the collapse of a cosmological void into a sheet. The void is located at the ridge between two larger voids (void-in-cloud scenario) . As the adjacent voids expand the highlighted void contracts into a thin sheet. Note how as the void radius decreases also its degree decreases (Fig. \ref{fig:collapsing_void-B}). 

We may be able to find observational evidence of the primordial void form which the Local Sheet was formed. In particular the existence of a population of luminous galaxies off the plane of the local wall \citep{Peebles10}. If our local wall was formed by the collapse of a small void one would expect galaxies at opposite walls of the collapsing void to pass through the newly formed wall \citep{Benitez13}. This scenario is plausible given our position at the edge of a large supercluster in which the Milky Way and its surrounding voids are embedded inside a large shallow overdensity \citep{Tully19}. 

We have, in fact, a dramatic example of a collapsing void in our own cosmic backyard. Just opposite to the local void there is a filamentary association of galaxies on the farther side of the ``southern" Leo Void: the Leo Spur \citep{Tully08b, Karachentsev15},  located at a distance of $\sim 7$ Mpc from our galaxy and  approaching to us as a whole with a radial velocity of $\sim 200$ km s$^{-1}$. This ``local velocity anomaly" can not be fully explained by the nearby massive structures \citep{Tully08b,Tully08}. It is, however, consistent with the void-in-cloud scenario described by \citet{Sheth04}. The southern void with a radius of $\sim 3.5$ lies well below the critical radius $R_{\textrm{\tiny{crit}}} \sim 9$ $h^{-1}$ Mpc  defining a collapsing void (see Fig. \ref{fig:von_neumann}).

%
\subsection{The Void network as a standard ruler}

In general cellular scaling relations assume an Euclidean geometry, however other space metrics are possible. The derivation of the von Neumann relation assumes a flat geometry where the integral of the mean curvature $k$ around a cell is $2\pi$. \citet{Avron92} generalized the von Neumann relation to include 2D curved spaces as $dA/dt = K( (n-6) +  (3A)(\pi R^2))$, where $R$ is the radius of curvature. \citet{Roth12} found a departure from the euclidean von Neumann relation measured in a 2D froth placed on the surface of a spherical dome. This departure  was interpreted as the result of the positively curved space being able to accommodate larger angles than the flat space. 

In the case of the network of Voids we should expect a similar dependence on the space metric. The Lewis, Aboav and von Neumann relations directly reflect the metric of space by means of the void-connectivity and size and therefore can provide a novel standard ruler for cosmology. The Lewis and Aboav relations can be measured from galaxy catalogues with accurate galaxy distance estimations and enough spatial sampling to resolve individual voids. In order to measure the von Neumann relation we need actual physical distances (in contrast to redshift distances) which can complicate its measurement in addition to the expected low magnitude of the signal. The two-dimensional von Neumann law has a dependence on spatial curvature as $R^{-2}$. The radius of curvature of the Universe is given by $R = D_0 / \sqrt{\Omega -1}$, where  $D_0 = H_0^{-1} \simeq 3000 h^{-1}$Mpc and $\Omega$ is the total density of the Universe. Any effect of spatial curvature on the von Neumann law will be very small for reasonable values of $\Omega$. The simulations we present here can not account for non-euclidian geometries. More sophisticated simulations, better theoretical understanding and future massive 3D galaxy surveys with accurate distance estimates are required to apply this relations as an independent standard ruler for cosmology.

\section{Data Availability}\label{sec:data_availbility}
The data used in this work can be provided upon request to the author. The Lagrangian Tessellation code used to compute density fields is available at  https://github.com/miguel-aragon/Lagrangian-Sheet-Density-Estimator.

\section{acknowledgments}
This research was partly funded by the Betty and Gordon Moore foundation, a New Frontiers of Astronomy and Cosmology grant from the Templeton Foundation and the ``Programa de Apoyo a Proyectos de Investigaci\'{o}n e Innovaci\'{o}n Tecnol\'{o}gica'' grant DGAPA-PAPIIT IA102020. The author would like to thank Mark Neyrinck for stimulating discussions.

\bibliography{references} 
\bibliographystyle{mn2e}   

\appendix

%
\section{A collapsing void}

\begin{figure}
\includegraphics[width=0.48\textwidth]{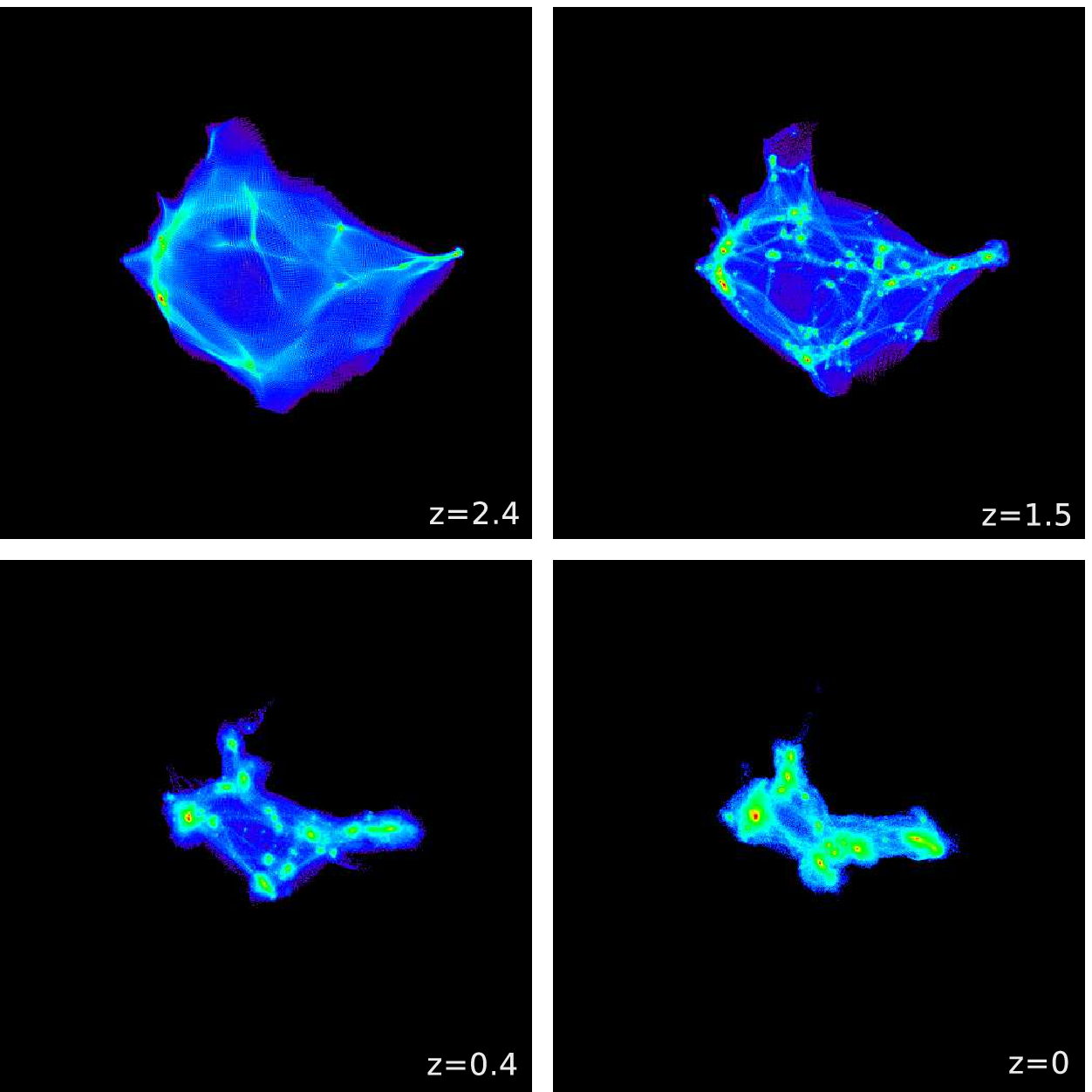}
\caption{Collapse of a void with radius $R \sim 9 h^{-1}$ Mpc at $z=10$.  Panels show the evolution of the void at $z=2.4, 1.5, 0.4$ and $0$. After collapsing, the void forms a wall  (here seen edge-on).}
 \label{fig:collapsing_void}
\end{figure}

\begin{figure}
\includegraphics[width=0.48\textwidth]{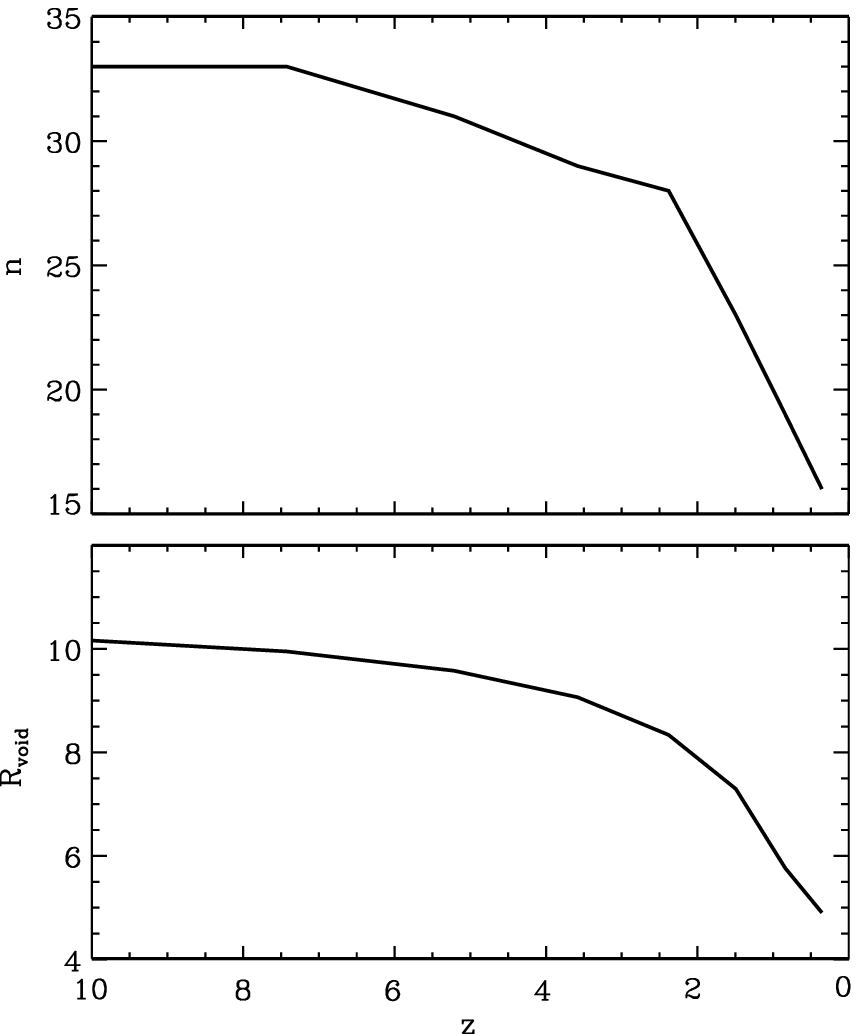}
\caption{Void degree (top) and effective radius (bottom) as a function of redshift for the void shown in Fig. \ref{fig:collapsing_void}. Void radius and degree were computed as described in Sec. \ref{sec:void_catalog} }
 \label{fig:collapsing_void-B}
\end{figure}

Figure \ref{fig:collapsing_void} shows the time evolution of a void  as it collapses from a starting radius of $\sim 9 h^{-1}$ Mpc to form a flat wall at the present time.  For visualization purposes we used a higher-resolution simulation than the one used in our analysis with $256^3$ particles and show the particles inside the void instead of the voxels. The radius and degree of the void as a function of redshift is shown in Fig. \ref{fig:collapsing_void-B}. It is interesting to note the abrupt decrease in both radius and degree starting around $z \sim 3$. See Fig. 4 in \citet{Sheth04} for another case of a collapsing void.

\end{document}